%% file: main.tex
\pdfoutput=1
\newif\ifAnon\Anonfalse
\newif\ifDraft\Draftfalse
\documentclass[sigconf,nonacm]{acmart}
\usepackage{tugraz_defaults}
\usepackage{enumitem}
\usepackage{subcaption}

\newcommand{\AttackPrimitive}{\emph{Fetch+\allowbreak Bounce}\xspace}
\newcommand{\SBLeak}{\emph{Data Bounce}\xspace}
\newcommand{\DataLeak}{\emph{Speculative Fetch+\allowbreak Bounce}\xspace}

\newcommand{\paragrabf}[1]{\paragraph{\textbf{#1}}}

\acmYear{2020}

\begin{document}

\date{}
 
\title[Store-to-Leak Forwarding]{Store-to-Leak Forwarding: Leaking Data on Meltdown-resistant CPUs (Updated and Extended Version)}

\author{Michael Schwarz}
\affiliation{\institution{CISPA Helmholtz Center for Information Security}}
\email{michael.schwarz@cispa.saarland}

\author{Claudio Canella}
\affiliation{\institution{Graz University of Technology}}
\email{claudio.canella@iaik.tugraz.at}

\author{Lukas Giner}
\affiliation{\institution{Graz University of Technology}}
\email{lukas.giner@iaik.tugraz.at}

\author{Daniel Gruss}
\affiliation{\institution{Graz University of Technology}}
\email{daniel.gruss@iaik.tugraz.at}

\renewcommand{\shortauthors}{Schwarz et al.}

\begin{abstract}
Meltdown and Spectre exploit microarchitectural changes the CPU makes during transient out-of-order execution.
Using side-channel techniques, these attacks enable leaking arbitrary data from memory. 
As state-of-the-art software mitigations for Meltdown may incur significant performance overheads, they are only seen as a temporary solution. 
Thus, software mitigations are disabled on more recent processors, which are not susceptible to Meltdown anymore.

In this paper, we show that Meltdown-like attacks are still possible on recent CPUs which are not vulnerable to the original Meltdown attack. 
We show that the store buffer---a microarchitectural optimization to reduce the latency for data stores---in combination with the TLB enables powerful attacks. 
We present several ASLR-related attacks, including a KASLR break from unprivileged applications, and breaking ASLR from JavaScript.
We can also mount side-channel attacks, breaking the atomicity of TSX, and monitoring control flow of the kernel.
Furthermore, when combined with a simple Spectre gadget, we can leak arbitrary data from memory. 
Our paper shows that Meltdown-like attacks are still possible, and software fixes are still necessary to ensure proper isolation between the kernel and user space.

This updated extended version of the original paper includes new results and explanations on the root cause of the vulnerability and shows how it is different to MDS attacks like Fallout~\cite{Canella2019Fallout}.
\end{abstract}



\maketitle

\newcommand{\inst}[1]{\texttt{#1}}

\section{Introduction} %
Modern processors have numerous optimizations to achieve the performance and efficiency that customers expect today.
Most of these optimizations, \eg CPU caches, are transparent for software developers and do not require changes in existing software. 
While the instruction-set architecture (ISA) describes the interface between software and hardware, it is only an abstraction layer for the CPUs microarchitecture. 
On the microarchitectural level, the CPU can apply any performance optimization as long as it does not violate the guarantees given by the ISA. 
Such optimizations also include pipelining or speculative execution. 
As the microarchitectural level is transparent and the optimizations are performed automatically, such optimizations are usually not or only sparsely documented. 
Furthermore, the main focus of microarchitectural optimizations is performance and efficiency, resulting in fewer security considerations than on the architectural level. 

In recent years, we have seen several attacks on the microarchitectural state of CPUs, making the internal state of the CPU visible~\cite{Percival2005,Osvik2006,Yarom2014Flush,Gruss2016Flush,Evtyushkin2018}. 
With knowledge about the internal CPU state, it is possible to attack cryptographic algorithms~\cite{Osvik2006,Percival2005,Yarom2014Flush,Irazoqui2014,Benger2014,Liu2015Last,Irazoqui2015SA}, spy on user interactions~\cite{Gruss2015Template,Lipp2016,Schwarz2018KeyDrown}, or covertly transmit data~\cite{Xu2011,Wu2014,Liu2015Last,Maurice2017Hello}. 
With the recent discovery of Meltdown~\cite{Lipp2018meltdown}, Foreshadow~\cite{Vanbulck2018foreshadow}, and Foreshadow-NG~\cite{Weisse2018foreshadow}, microarchitectural attacks advanced to a state where not only metadata but arbitrary data can be leaked.
These attacks exploit the property that many CPUs still continue working out-of-order with data even if the data triggered a fault when loading it, \eg due to a failed privilege check. 
Although the data is never architecturally visible, it can be encoded into the microarchitectural state and made visible using microarchitectural side-channel attacks. 

While protecting against side-channel attacks was often seen as the duty of developers~\cite{Bernstein2005,Irazoqui2014}, Meltdown and Foreshadow-NG showed that this is not always possible. 
These vulnerabilities, which are present in most Intel CPUs, break the hardware-enforced isolation between untrusted user applications and the trusted kernel. 
Hence, these attacks allow an attacker to read arbitrary memory, against which a single application cannot protect itself. 

As these CPU vulnerabilities are deeply rooted in the CPU, close to or in the critical path, they cannot be fixed with microcode updates, but the issue is fixed on more recent processors~\cite{IntelSpecAnalysis,IntelMitigations,Cutress2018Spectre}. 
Due to the severity of these vulnerabilities, and the ease to exploit them, all major operating systems rolled out software mitigations to prevent exploitation of Meltdown~\cite{Gruss2018Kernel,Hansen2017kaiser,KVAShadow2018,OSXMeltdown2018}.
The software mitigations are based on the idea of separating user and kernel space in stricter ways~\cite{Gruss2017KASLR}.
While such a stricter separation does not only prevent Meltdown, it also prevents other microarchitectural attacks on the kernel~\cite{Gruss2017KASLR}, \eg microarchitectural KASLR (kernel address-space layout randomization) breaks~\cite{Hund2013,Gruss2016Prefetch,Jang2016}. 
Still, software mitigations may incur significant performance overheads, especially for workloads that require frequent switching between kernel and user space~\cite{Gruss2018Kernel}. 
Thus, CPU manufacturers solved the root issue directly in hardware, making the software mitigations obsolete. 

Although new CPUs are not vulnerable to the original Meltdown attack, we show that similar Meltdown-like effects can still be observed on such CPUs. 
In this paper, we investigate the store buffer and its microarchitectural side effects. 
The store buffer is a microarchitectural element which serializes the stream of stores and hides the latency when storing values to memory. 
It works similarly to a queue, completing all memory stores asynchronously while allowing the CPU to continue executing the execution stream out of order. 
To guarantee the consistency of subsequent load operations, load operations have to first check the store buffer for pending stores to the same address. 
If there is a store-buffer entry with a matching address, the load is served from the store buffer. 
This so-called store-to-load forwarding has been exploited in Spectre v4~\cite{Horn2018spectre4}, where the load and store go to different virtual addresses mapping the same memory location.
Consequently, the virtual address of the load is not found in the store buffer and a stale value is read from the caches or memory instead. 
However, due to the asynchronous nature of the store buffer, Meltdown-like effects are visible, as store-to-load forwarding also happens after an illegal memory store. 

We focus on correct store-to-load forwarding, \ie no false dependencies.
We present three basic attack techniques that each leak side-channel information from correct store-to-load forwarding.
First, \SBLeak, which exploits that stores to memory are forwarded even if the target address of the store is inaccessible to the user, \eg kernel addresses.
With \SBLeak we break KASLR, reveal the address space of Intel SGX enclaves, and even break ASLR from JavaScript.
Second, \AttackPrimitive, which combines \SBLeak with the TLB side channel.
With \AttackPrimitive we monitor kernel activity on a page-level granularity.
Third, \DataLeak, which combines \AttackPrimitive with speculative execution, leading to arbitrary data leakage from memory.
\DataLeak does not require shared memory between the user space and the kernel~\cite{Kocher2019}, and the leaked data is not encoded in the cache.
Hence, \DataLeak even works with countermeasures in place which only prevent cache covert channels. %

We conclude that the hardware fixes for Meltdown are not sufficient on new CPUs. 
We stress that due to microarchitectural optimizations, security guarantees for isolating the user space from kernel space are not as strong as they should be. 
Therefore, we highlight the importance of keeping the already deployed additional software-based isolation of user and kernel space~\cite{Gruss2017KASLR}. 

\paragraph{Contributions.}
The contributions of this work are:
\begin{enumerate}
	\item We discover a Meltdown-like effect around the store buffer on Intel CPUs (\SBLeak).
	\item We present \AttackPrimitive, a side-channel attack leveraging the store buffer and the TLB.
	\item We present a KASLR break, and an ASLR break from both JavaScript and SGX, and a covert channel. 
	\item We show that an attacker can still leak kernel data even on CPUs where Meltdown is fixed (\DataLeak). %
	\item We provide an analysis of the microarchitecture explaining the root cause of the vulnerability we discovered.
\end{enumerate}

\paragraph{Outline.}
\cref{sec:background} provides background on transient execution attacks.
We describe the basic effects and attack primitives in \cref{sec:attackprimitives}.
We present KASLR, and ASLR breaks with \SBLeak in \cref{sec:attack-aslr}.
We show how control flow can be leaked with \AttackPrimitive in \cref{sec:attack-cf}.
We demonstrate how \DataLeak allows leaking kernel memory on fully patched hardware and software in \cref{sec:attack-leak}.
We provide a root-cause analysis on the microarchitectural level in \cref{sec:analysis}.
We discuss the context of our attack and related work in \cref{sec:discussion}.
We conclude in \cref{sec:conclusion}.

\paragraph{Responsible Disclosure.}
We responsibly disclosed our initial research to Intel on January 18, 2019. 
Intel verified our findings. 
The findings were part of an embargo ending on May 14, 2019.

\section{Background}%
\label{sec:background}
In this section, we describe the background required for this paper. 
We give a brief overview of caches, transient execution and transient execution attacks, store buffers, virtual memory, and Intel SGX. 

\subsection{Cache Attacks}\label{sec:background:caches}
Processor speeds increased massively over the past decades.
While the bandwidth of modern main memory (DRAM) has increased accordingly, the latency has not decreased to the same extent.
Consequently, it is essential for the processor to fetch data from DRAM ahead of time and buffer it in faster internal storage.
For this purpose, processors contain small memory buffers, called caches, that store frequently or recently accessed data.
In modern processors, the cache is organized in a hierarchy of multiple levels, with the lowest level being the smallest but also the fastest.
In each subsequent level, the size and access time increases.

As caches are used to hide the latency of memory accesses, they inherently introduce a timing side channel.
Many different cache attack techniques have been proposed over the past two decades~\cite{Kocher1996,Bernstein2005,Osvik2006,Yarom2014Flush,Gruss2016Flush}.
Today, the most important techniques are \PrimeProbe~\cite{Osvik2006, Percival2005} and \FlushReload~\cite{Yarom2014Flush}.
Variants of these attacks that are used today exploit that the last-level cache is shared and inclusive on many processors.
\PrimeProbe attacks constantly measure how long it takes to fill an entire cache set.
Whenever a victim process accesses a cache line in this cache set, the measured time will be slightly higher.
In a \FlushReload attack, the attacker constantly flushes the targeted memory location using the \clflush instruction.
The attacker then measures how long it takes to reload the data.
Based on the reload time the attacker determines whether a victim has accessed the data in the meantime.
Due to its fine granularity, \FlushReload has been used for attacks on various computations, \eg web server function calls~\cite{Zhang2014}, user input~\cite{Gruss2015Template,Lipp2016, Schwarz2018KeyDrown}, kernel addressing information~\cite{Gruss2016Prefetch}, and cryptographic algorithms~\cite{Yarom2014Flush, Irazoqui2014, Benger2014}.

Covert channels are a particular use case of side channels.
In this scenario, the attacker controls both the sender and the receiver and tries to leak information from one security domain to another, bypassing isolation imposed on the functional or the system level.
\FlushReload as well as \PrimeProbe have both been used in high-performance covert channels~\cite{Liu2015Last,Maurice2017Hello,Gruss2016Flush}.

\subsection{Transient-execution Attacks}
Modern processors are highly complex and large systems.
Program code has a strict in-order instruction stream.
However, if the processor would process this instruction stream strictly in order, the processor would have to stall until all operands of the current instruction are available, even though subsequent instructions might be ready to run.
To optimize this case, modern processors first fetch and decode an instruction in the frontend.
In many cases, instructions are split up into smaller micro-operations (\muops)~\cite{Fog2016}.
These \muops are then placed in the so-called Re-Order Buffer (ROB).
\muops that have operands also need storage space for these operands.
When a \muop is placed in the ROB, this storage space is dynamically allocated from the load buffer, for memory loads, the store buffer, for memory stores, and the register file, for register operations.
The ROB entry only references the load buffer and store buffer entries.
While the operands of a \muop still might not be available after it was placed in the ROB, we can now schedule subsequent \muops in the meantime.
When a \muop is ready to be executed, the scheduler schedules them for execution.
The results of the execution are placed in the corresponding registers, load buffer entries, or store buffer entries.
When the next \muop in order is marked as finished, it is retired, and the buffered results are committed and become architectural.

As software is rarely purely linear, the processor has to either stall execution until a (conditional) branch is resolved or speculate on the most likely outcome and start executing along the predicted path.
The results of those predicted instructions are placed in the ROB until the prediction has been verified.
In the case where the prediction was correct, the instructions are retired in order.
If the prediction was wrong, the processor reverts all architectural changes, flushes the pipeline and the ROB but does not revert any microarchitectural state changes, \ie loading data into the cache or TLB.
Similarly, when an interrupt occurs, operations already executed out of order must be flushed from the ROB.
We refer to instructions that have been executed speculatively or out-of-order but were never committed as \textit{transient instructions}~\cite{Kocher2019, Lipp2018meltdown, Canella2019A}.
Spectre-type attacks~\cite{Kocher2019,Kiriansky2018speculative,Maisuradze2018spectre5,Koruyeh2018spectre5,Horn2018spectre4,Chen2018SGXpectre,Canella2019A} exploit the transient execution of instructions before the misprediction by one of the processor's prediction mechanisms is discovered.
Meltdown-type attacks~\cite{Lipp2018meltdown,Vanbulck2018foreshadow, Kiriansky2018speculative,Weisse2018foreshadow,Stecklina2018,ARMSpecAnalysis,IntelSpecAnalysis,IntelMitigations,Canella2019A} exploit the transient execution of instructions before an interrupt or fault is handled.

\subsection{Store Buffer}\label{sec:primitive-store-buffer}
To interact with the memory subsystem (and also to hide some of the latency), modern CPUs have store and load buffers (also called memory order buffer~\cite{Islam2019Spoiler}) which act as a queue. 
The basic mechanism is that the load buffer contains requests for data fetches from memory, while the store buffer contains requests for data writes to memory. 

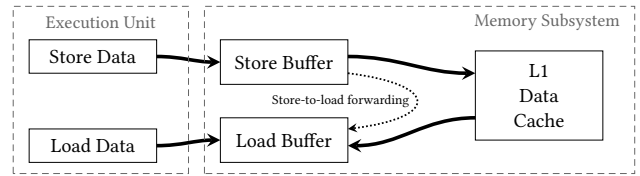
\begin{figure}[t]
\resizebox{\hsize}{!}{
 \input{img/store_buffer.tikz}
}
 \caption{A store operation stores the data in the store buffer before it is written to the L1 data cache. 
Subsequent loads can be satisfied from the store buffer if the data is not yet in the L1 data cache.
This is called store-to-load forwarding.}
 \label{fig:store-buffer}
\end{figure}

As long as the store buffer is not exhausted, memory stores are simply enqueued to the store buffer in the order they appear in the execution stream, \ie directly linked to a ROB entry. 
This allows the CPU to continue executing instructions from the current execution stream, without having to wait for the actual write to finish. 
This optimization makes sense, as writes in many cases do not influence subsequent instructions, \ie only loads to the same address are affected. 
Meanwhile, the store buffer asynchronously processes the stores, ensuring that the stores are written to memory. 
Thus, the store buffer avoids that the CPU has to stall while waiting for the memory subsystem to finish the write.
At the same time, it guarantees that writes reach the memory subsystem in order, despite out-of-order execution. 
\Cref{fig:store-buffer} illustrates the role of the store buffer, as a queue between the store-data execution unit and the memory subsystem, \ie the L1 data cache. 

For every store operation that is added to the ROB, an entry is allocated in the store buffer. 
This entry requires both the virtual and physical address of the target. 
Only if there is no free entry in the store buffer, the frontend stalls until there is an empty slot available in the store buffer again~\cite{Intel_opt}. 
Otherwise, the CPU can immediately continue adding subsequent instructions to the ROB and execute them out of order. 
On Intel CPUs, the store buffer has up to 56 entries~\cite{Intel_opt}.

According to Intel patents, the store buffer consists of two separate buffers: the \emph{Store Address Buffer} and the \emph{Store Data Buffer}~\cite{abramson2002method,Abramson1996}.
The store instruction is decoded into two \muops, one for storing the address and one for storing data.
Those two instructions can execute in either order, depending on which is ready first.

Although the store buffer hides the latency of stores, it also increases the complexity of loads. 
Every load has to search the store buffer for pending stores to the same address in parallel to the regular L1 lookup. 
If the address of a load matches the address of a preceding store, the value can be directly used from the store-buffer entry. 
This optimization for subsequent loads is called store-to-load forwarding~\cite{Hooker2013STL}. 

Depending on the implementation of the store buffer, there are various ways of implementing such a search required for store-to-load forwarding, \eg using content-addressable memory~\cite{Wong2014STL}.
As loads and stores on x86 do not have to be aligned, a load can also be a partial match of a preceding store. 
Such a load with a partial match of a store-buffer entry can either stall, continue with stale data, or be resolved by the CPU by combining values from the store buffer and the memory~\cite{Wong2014STL}.

Moreover, to speed up execution, the CPU might wrongly predict that values should be fetched from memory although there was a previous store, but the target of the previous store is not yet resolved. 
As a result, the processor can continue transient execution with wrong values, \ie stale values from memory instead of the recently stored value. 
This type of misprediction was exploited in Spectre v4 (Speculative Store Bypass)~\cite{Horn2018spectre4}, also named Spectre-STL~\cite{Canella2019A}. 

To speed up store-to-load forwarding, the processor might speculate that a load matches the address of a subsequent store if only the least significant 12 bits match~\cite{Wong2014STL}.
This performance optimization can further reduce the latency of loads, but also leaks information across hyperthreads~\cite{Sullivan2018minefields}.
Furthermore, a similar effect also exists if the least significant 20 bits match~\cite{Islam2019Spoiler}.
If the load causes a fault, this even leads to a Meltdown-type data leakage~\cite{Canella2019Fallout}. 

\subsection{Address Translation}
Memory isolation is the basis of modern operating system security.
For this purpose, processors support virtual memory as an abstraction and isolation mechanism.
Processes work on virtual addresses instead of physical addresses and can architecturally not interfere with each other unintentionally, as the virtual address spaces are largely non-overlapping.
The processor translates virtual addresses to physical addresses through a multi-level page translation table.
The location of the translation table is indicated by a dedicated register, \eg CR3 on Intel architectures.
The operating system updates the register upon context switch with the physical address of the top-level translation table of the next process.
The translation table entries keep track of various properties of the virtual memory region, \eg user-accessible, read-only, non-executable, and present.

\paragrabf{Translation Lookaside Buffer (TLB)}
The translation of a virtual to a physical address is time-consuming as the translation tables are stored in physical memory.
On modern processors, the translation is required even for L1 cache accesses.
Hence, the translation must be faster than the full L1 access, \eg 4 cycles on recent Intel processors.
Caching translation tables in regular data caches~\cite{Intel_vol3} is not sufficient.
Therefore, processors have smaller special caches, translation-lookaside buffers (TLBs) to cache page table entries.

\subsection{Address Space Layout Randomization}\label{sec:background:aslr}
To exploit a memory corruption bug, an attacker often requires knowledge of addresses of specific data.
To impede such attacks, different techniques like address space layout randomization (ASLR), non-executable stacks, and stack canaries have been developed.
KASLR extends ASLR to the kernel, randomizing the offsets where code, data, drivers, and other mappings are located on every boot.
The attacker then has to guess the location of (kernel) data structures, making attacks harder.

The double page fault attack by Hund~\etal\cite{Hund2013} breaks KASLR.
An unprivileged attacker accesses a kernel memory location and triggers a page fault.
The operating system handles the page fault interrupt and hands control back to an error handler in the user program.
The attacker now measures how much time passed since triggering the page fault.
Even though the kernel address is inaccessible to the user, the address translation entries are copied into the TLB.
The attacker now repeats the attack steps, measuring the execution time of a second page fault to the same address.
If the memory location is valid, the handling of the second page fault will take less time as the translation is cached in the TLB.
Thus, the attacker learns whether a memory location is valid even though the address is inaccessible to user space.

The same effect has been exploited by Jang~\etal\cite{Jang2016} in combination with Intel TSX.
Intel TSX extends the x86 instruction set with support for hardware transactional memory via so-called TSX transactions.
A TSX transaction is aborted without any operating system interaction if a page fault occurs within it.
This reduces the noise in the timing differences that was present in the attack by Hund~\etal\cite{Hund2013} as the page fault handling of the operating system is skipped.
Thus, the attacker learns whether a kernel memory location is valid with almost no noise at all.

The prefetch side channel presented by Gruss~\etal\cite{Gruss2016Prefetch} exploits the software prefetch instruction.
The execution time of the instruction is dependent on the translation cache that holds the right entry.
Thus, the attacker not only learns whether an inaccessible address is valid but also the corresponding page size.

\subsection{Intel SGX}
As computer usage has changed over the past decades, the need for a protected and trusted execution mechanism has developed.
To protect trusted code, Intel introduced an instruction-set extension starting with the Skylake microarchitecture, called Software Guard Extension (SGX)~\cite{Intel_vol3}.
SGX splits applications into two code parts, a trusted and an untrusted part.
The trusted part is executed within a hardware-backed enclave.
The processor guarantees that memory belonging to the enclave cannot be accessed by anyone except the enclave itself, not even the operating system.
The memory is encrypted and, thus, also cannot be read directly from the DRAM module.
Beyond this, there is no virtual memory isolation between trusted and untrusted part.
Consequently, the threat model of SGX assumes that the operating system, other applications, and even the remaining hardware might be compromised or malicious.
However, memory-safety violations~\cite{Lee2017SGXROP}, race conditions~\cite{Weichbrodt2016}, or side channels~\cite{Brasser2017sgx,Schwarz2017SGX} are considered out of scope.

\begin{figure}[t]
 \centering
 \resizebox{0.9\hsize}{!}{
 \input{img/sgx.tikz} 
 }
 \caption{In the SGX model, applications consist of an untrusted host application and a trusted enclave. 
The hardware prevents any direct access to the enclave code or data. 
 The untrusted part uses the \texttt{EENTER} instruction to call enclave functions that are exposed by the enclave.
 }
 \label{fig:sgx}
\end{figure}
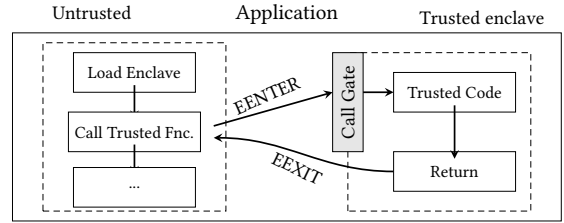

The untrusted part can only enter the enclave through a defined interface which is conceptually similar to system calls.
After the trusted execution, the result of the computation, as well as the control flow, is handed back to the calling application.
The process of invoking a trusted enclave function is illustrated in \cref{fig:sgx}.
To enable out-of-the-box data sharing capabilities, the enclave has full access to the entire address space of the host.
As this protection is not symmetric, it gives rise to enclave malware~\cite{Schwarz2019SGXMalware}.

\section{Attack Primitives}\label{sec:attackprimitives}%
In this section, we introduce the three basic mechanisms for our attacks.
First, \SBLeak, which exploits that stores to memory are forwarded even if the target address of the store is inaccessible to the user.
We use \SBLeak to break both user and kernel space ASLR (\cf \Cref{sec:attack-aslr}).
Second, we exploit interactions between \SBLeak and the TLB in \AttackPrimitive.
\AttackPrimitive enables attacks on the kernel on a page-level granularity, similar to controlled-channel attacks~\cite{Xu2015controlled}, page-cache attacks~\cite{Gruss2019page}, TLBleed~\cite{Gras2018}, and DRAMA~\cite{Pessl2016} (\cf \Cref{sec:attack-cf}).
Third, we augment \AttackPrimitive with speculative execution in \DataLeak.
\DataLeak leads to arbitrary data leakage from memory (\cf \Cref{sec:attack-leak}). 

As described in \cref{sec:primitive-store-buffer}, unsuccessful or incorrect address matching in the store-to-load forwarding implementation can enable different attacks.
For our attacks, we focus solely on the case where the address matching in the store-to-load forwarding implementation is successful and correct.
We exploit store-to-load forwarding in the case where the address of the store and load are \emph{exactly the same}, \ie we do not rely on any misprediction or aliasing effects.

\subsection{\SBLeak}
Our first attack primitive, \SBLeak, exploits the property of the store buffer that the full physical address is required for a valid entry. 
Although the store-buffer entry is already reserved in the ROB, the actual store can only be forwarded if the virtual and physical address of the store target are known~\cite{Intel_opt}.

Thus, stores can only be forwarded if the physical address of the store target can be resolved. 
As a consequence, virtual addresses without a valid mapping to physical addresses cannot be forwarded to subsequent loads. 
The basic idea of \SBLeak is to check whether a data write is forwarded to a data load from the same address. 
If the store-to-load forwarding is successful for a chosen address, we know that the chosen address can be resolved to a physical address. 
If done na\"ively, such a test would destroy the currently stored value at the chosen address due to the write if the address is writable. 
Thus, we only test the store-to-load forwarding for an address in the transient-execution domain, \ie the write is never committed architecturally. 

\begin{figure}[t]
 \begin{tikzpicture}
  \node[anchor=west] at (0.25, 1.5) {\circleds{1} \quad \texttt{mov (0) $\to$ \$dummy}};
  \node[anchor=west] at (0.25, 1) {\circleds{2} \quad \texttt{mov \$x $\to$ (p)}};
  \node[anchor=west] at (0.25, 0.5) {\circleds{3} \quad \texttt{mov (p) $\to$ \$value}};
  \node[anchor=west] at (0.25, 0) {\circleds{4} \quad \texttt{mov (\$mem + \$value * 4096) $\to$ \$dummy}};
 \end{tikzpicture}

 \caption{\SBLeak writes a known value to an accessible or inaccessible memory location, reads it back, encodes it into the cache, and finally recovers the value using a \FlushReload attack. If the recovered value matches the known value, the address is backed by a physical page.}\label{fig:sbleak}
\end{figure}

\Cref{fig:sbleak} illustrates the basic principle of \SBLeak. 
First, we start transient execution. 
The easiest way is by generating a fault (\circleds{1}) and catching it (\eg with a signal handler) or suppressing it (\eg using Intel TSX). 
Alternatively, transient execution can be induced through speculative execution using a misspeculated branch~\cite{Kocher2019}, call~\cite{Kocher2019}, or return~\cite{Maisuradze2018spectre5,Koruyeh2018spectre5}. 
For a chosen address $p$, we store any chosen value $x$ using a simple data store operation (\circleds{2}). 
Subsequently, we read the value stored at address $p$ (\circleds{3}) and encode it in the cache (\circleds{4}) in the same way as with Meltdown~\cite{Lipp2018meltdown}. 
That is, depending on the value read from $p$, we access a different page of the contiguous memory $mem$, resulting in the respective page being cached.
Using a straightforward \FlushReload attack on the \SIx{256} pages of $mem$, the page with the lowest access time (\ie the cached page) directly reveals the value read from $p$. 

We can then distinguish two different cases as follows.

\begin{description}
\item[Store-to-load forwarding.]
If the value read from $p$ is $x$, \ie, the $x$-th page of $mem$ is cached, the store was forwarded to the load. 
Thus, we know that $p$ is backed by a physical page. 
The choice of the value $x$ is of no importance for \SBLeak. 
Even in the unlikely case that $p$ already contains the value $x$ and the CPU reads the stale value from memory instead of the previously stored value $x$, we still know that $p$ is backed by a physical page.

\item[No store-to-load forwarding.]
If no page of $mem$ is cached, the store was not forwarded to the subsequent load. 
This can have either a temporary reason or a permanent reason. 
If the virtual address is not backed by a physical page, the store-to-load forwarding always fails, \ie even retrying the experiment will not be successful. 
Different reasons to not read the written value back are, \eg interrupts (context switches, hardware interrupts) or errors in distinguishing cache hits from cache misses (\eg due to power scaling). 
However, we found that if \SBLeak fails multiple times when repeated for $addr$, it is almost certain that $addr$ is not backed by a physical page. 
\end{description}

In summary, if a value ``bounces back'' from a virtual address, the virtual address must be backed by a physical page. 
This effect can be exploited within the virtual address space of a process, \eg to break ASLR in a sandbox (\cf \Cref{sec:attack-aslr-js}). 
On CPUs where Meltdown is mitigated in hardware, KAISER~\cite{Gruss2017KASLR} is not enabled, and the kernel is again mapped in user space~\cite{Cutress2018Spectre}. 
In this case, we can also apply \SBLeak to kernel addresses. 
Even though we cannot \textit{access} the data stored at the kernel address, we are still able to detect whether a particular kernel address is backed by a physical page. 
Thus, \SBLeak can still be used to break KASLR (\cf \Cref{sec:attack-aslr-kernel}) on processors with in-silicon patches against Meltdown.

\subsection{\AttackPrimitive}

Our second attack primitive, \AttackPrimitive, augments \SBLeak with an additional interaction effect of the TLB and the store buffer. 
With this combination, it is also possible to detect the recent usage of physical pages. 

\SBLeak is a very reliable side channel, making it is easy to distinguish valid from invalid addresses, \ie whether a virtual page is backed by a physical page. 
Additionally, the success rate (\ie how often \SBLeak has to be repeated) for valid addresses directly depends on which translations are stored in the TLB. 
With \AttackPrimitive, we further exploit this TLB-related side-channel information by analyzing the success rate of \SBLeak. 

The store buffer requires the physical address of the store target (\cf \Cref{sec:primitive-store-buffer}). 
If the translation from virtual to physical address for the target address is not cached in the TLB, the store triggers a page-table walk to resolve the physical address. 
On our test machines, we observed that in this case the store-to-load forwarding fails once, \ie as the physical address of the store is not known, it is not forwarded to the subsequent load.
In the other case, when the physical address is already known to the TLB, the store-to-load forwarding succeeds immediately. 

\begin{figure}[t]
 \begin{tikzpicture}
  \node[anchor=west] at (0.25, 1.5) {\circleds{1} \quad \texttt{\textbf{for} retry = 0...2}};
  \node[anchor=west] at (0.25, 1) {\phantom{\circleds{2}} \qquad \texttt{mov \$x $\to$ (p)}};
  \node[anchor=west] at (0.25, 0.5) {\circleds{2} \qquad \texttt{mov (p) $\to$ \$value}};
  \node[anchor=west] at (0.25, 0) {\phantom{\circleds{4}} \qquad \texttt{mov (\$mem + \$value * 4096) $\to$ \$dummy}};
  \node[anchor=west] at (0.25, -0.5) {\circleds{3} \qquad \texttt{\textbf{if} flush\_reload(\$mem + \$x * 4096) \textbf{then} break}};
  
 \end{tikzpicture}

 \caption{\AttackPrimitive repeatedly executes \SBLeak. If \SBLeak is successful on the first try, the address is in the TLB. If it succeeds on the second try, the address is valid but not in the TLB.}\label{fig:sbleak-tlb}
\end{figure}

\begin{figure}[t]
 \input{img/sb-tlb.tikz}
 \caption{Mounting \AttackPrimitive on a virtual memory range allows to clearly distinguish mapped from unmapped addresses. Furthermore, for every page, it allows to distinguish whether the address translation is cached in the TLB.}\label{fig:sbleak-tlb-plot}
\end{figure}
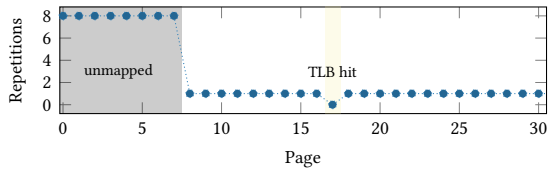

With \AttackPrimitive, we exploit that \SBLeak succeeds immediately if the mapping for this address is already cached in the TLB. 
\Cref{fig:sbleak-tlb} shows how \AttackPrimitive works. 
The basic idea is to repeat \SBLeak (\circleds{2}) multiple times (\circleds{1}). 
There are 3 possible scenarios, which are also illustrated in \Cref{fig:sbleak-tlb-plot}. 
\begin{description}
 \item[TLB Hit.]
If the address of the store is in the TLB, \SBLeak succeeds immediately, and the loop is aborted (\circleds{3}). 
Thus, \texttt{retry} is 0 after the loop.
 
 \item[TLB Miss.]
If the address of the store is not in the TLB, \SBLeak fails in the first attempt, as the physical address needs to be resolved before store-to-load forwarding. 
However, in this case, \SBLeak succeeds in the second attempt (\ie, \texttt{retry} is 1). 
 
 \item[Invalid Address.]
If the address is invalid, \texttt{retry} is larger than 1.
As only valid address are stored in the TLB~\cite{Jang2016}, and the store buffer requires a valid physical address, store-to-load forwarding can never succeed. 
The higher \texttt{retry}, the (exponentially) more confidence is gained that the address is indeed not valid.
\end{description}

As \SBLeak can be used on inaccessible addresses (\eg kernel addresses), this also works for \AttackPrimitive. 
Hence, with \AttackPrimitive it is possible to deduce for any virtual address whether it is currently cached in the TLB. 
The only requirement for the virtual address is that it is mapped to the attacker's address space. 

\AttackPrimitive is not limited to the data TLB (dTLB), but can also leak information from the instruction TLB (iTLB). 
Thus, in addition to recent data accesses, it is also possible to detect which code pages have been executed recently.
Again, this also works for inaccessible addresses, \eg kernel memory. 

Moreover, \AttackPrimitive cannot only be used to check whether a (possibly) inaccessible address is in the TLB but also force such an address into the TLB. 
While this effect might be exploitable on its own, we do not further investigate this side effect. 
For a real-world attack (\cf \Cref{sec:attack-cf}) this is an undesired side effect, as every measurement with \AttackPrimitive destroys the information. 
Thus, to repeat \AttackPrimitive for one address, we must evict the TLB in between, \eg using the strategy proposed by Gras~\etal\cite{Gras2018}. 

\subsection{\DataLeak}

Our third attack primitive, \DataLeak, augments \AttackPrimitive with transient-execution side effects on the TLB. 
The TLB is also updated during transient execution~\cite{Schwarz2019netspectre}.
That is, we can even observe \emph{transient} memory accesses with \AttackPrimitive. 

\begin{figure}[t]
\resizebox{\hsize}{!}{
    \input{img/spec-leak.tikz}
}
 \caption{\DataLeak allows an attacker to use Spectre gadgets to leak data from the kernel, by encoding them in the TLB. The advantage over regular Spectre attacks is that no shared memory is required, gadgets are simpler as an attacker does not require control of the array base address but only over \texttt{x}. All cache-based countermeasures are circumvented.}\label{fig:spectre-tlb}
\end{figure}
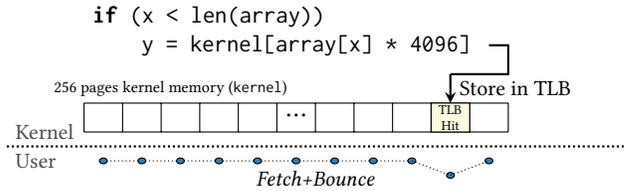

As a consequence, \DataLeak is a novel way to exploit Spectre. 
Instead of using the cache as a covert channel in a Spectre attack, we leverage the TLB to encode the leaked data. 
The advantage of \DataLeak over the original Spectre attack is that there is no requirement for shared memory between user and kernel space.
The attacker only needs control over \texttt{x} to leak arbitrary memory contents from the kernel.
\Cref{fig:spectre-tlb} illustrates the encoding of the data, which is similar to the original Spectre attack~\cite{Kocher2019}. 
Depending on the value of the byte to leak, we access one out of 256 pages. 
Then, \AttackPrimitive is used to detect which of the pages has a valid translation cached in the TLB. 
The cached TLB entry directly reveals the leaked byte.

\subsection{Performance and Accuracy}

All of the 3 attack primitives work on a page-level granularity, \ie \SI{4}{\kilo B}. 
This limit is imposed by the underlying architecture, as virtual-to-physical mappings can only be specified at page granularity. 
Consequently, TLB entries also have a page granularity. 
Hence, the spatial accuracy is the \SI{4}{\kilo B} page size, which is the same on all CPUs. 

While the \textit{spatial} accuracy is always the same, the \textit{temporal} accuracy varies between microarchitectures and implementations. 
On the i9-9900K, we measured the time of \SBLeak over \SIx{1000000} repetitions and it takes on average 560 cycles per execution. 
In this implementation, we use Intel TSX to suppress the exception. 
When resorting to a signal handler for catching exceptions instead of Intel TSX, one execution of \SBLeak takes on average \SIx{2300} cycles. 

\AttackPrimitive and \DataLeak do not require any additional active part in the attack. 
They execute \SBLeak 3 times, thus the execution time is exactly three times higher than the execution time of \SBLeak. 

\SBLeak has the huge advantage that there are no false positives. 
Store-to-load forwarding does not work for invalid addresses. 
Additionally, \FlushReload when applied to individual pages does not have false positives, as the prefetcher on Intel CPUs cannot cross page boundaries~\cite{Intel_vol3}. 
Thus, if a virtual address is not backed by a physical page, \SBLeak never reports this page as mapped. 

The number of false negatives reported by \SBLeak is also negligible. 
We do not exploit any race condition~\cite{Lipp2018meltdown,Vanbulck2018foreshadow} or aliasing effects~\cite{Sullivan2018minefields} but rather a missing permission check. 
Hence, the store-to-load forwarding works reliably as expected~\cite{Wong2014STL}. 
This can also be seen in the real-world attacks (\cf \Cref{sec:attack-aslr}), where the F1-score, \ie the harmonic average of precision and recall, is almost always perfect. 
We can conclude that \SBLeak is a highly practical side-channel attack with perfect precision and recall.

\subsection{Environments}

\begin{table}
\caption{Environments where we evaluated \SBLeak, \AttackPrimitive, and \DataLeak.}
\label{tab:setups}
\adjustbox{max width=\hsize}{
 \begin{tabular}{llccc}
  \rotatebox{30}{\hspace{-0.33em}{Environment}} & \rotatebox{30}{\hspace{-0.33em}{CPU}} & \rotatebox{30}{\hspace{-0.33em}{\SBLeak}} & \rotatebox{30}{\hspace{-0.33em}{\AttackPrimitive}} & \rotatebox{30}{\hspace{0.33em}{\parbox{2cm}{\centering Speculative\\\AttackPrimitive}}} \\ \hline
  Lab & Pentium 4 531 & \cmark & \xmark & \xmark \\
  Lab & i5-3230M & \cmark & \cmark & \cmark \\
  Lab & i7-4790 & \cmark & \cmark & \cmark \\
  Lab & i7-6600U & \cmark & \cmark & \cmark \\
  Lab & i7-6700K & \cmark & \cmark & \cmark \\
  Lab & i7-8565U & \cmark & \cmark & \cmark \\
  Lab & i7-8650U & \cmark & \cmark & \cmark \\
  Lab & i9-9900K & \cmark & \cmark & \cmark \\
  Lab & E5-1630 v4 & \cmark & \cmark & \cmark \\
  Lab & Xeon Silver 4208 & \xmark & \xmark & \xmark \\
  Cloud & E5-2650 v4 & \cmark & \cmark & \cmark \\
  Cloud & Unspecified Cascade Lake & \xmark & \xmark & \xmark \\
  \hline
 \end{tabular}
}
\end{table}

We evaluated \SBLeak, \AttackPrimitive, and \DataLeak on multiple Intel CPUs. 
All attack primitives worked on all tested CPUs up until Whiskey Lake and Coffee Lake R (both released end of 2018)\footnote{Since the time of writing, we have tested our attack primitives on even more CPUs and it appears that newer CPUs include fixes.}.
\SBLeak even works on Pentium 4 Prescott CPUs (released 2004). 
\Cref{tab:setups} contains the complete list of CPUs we used to evaluate the attacks. 

The primitives are not limited to the Intel Core microarchitecture, but also work on the Intel Xeon microarchitecture. 
Thus, these attacks are not limited to consumer devices, but can also be used in the cloud. 
Furthermore, the attack primitives even work on CPUs which have silicon fixes for Meltdown and Foreshadow, such as the i7-8565U and i9-9900K~\cite{Cutress2018Spectre}.  

For AMD, and ARM CPUs, we were not able to reproduce any of our attack primitives, limiting the attacks to Intel CPUs.

\section{Attacks on ASLR}\label{sec:attack-aslr}%
In this section, we evaluate our attack on ASLR in different scenarios.
As \SBLeak can reliably detect whether a virtual address is backed by a physical page, it is well suited for breaking all kinds of ASLR. 
In \Cref{sec:attack-aslr-kernel}, we show that \SBLeak is the fastest way and most reliable side-channel attack to break KASLR on Linux, and Windows, both in native environments as well as in virtual machines.\footnote{Since the time of writing, we discovered an even faster and more reliable KASLR break~\cite{Canella2020kaslr}.}
In \Cref{sec:attack-aslr-sgx}, we demonstrate that \SBLeak also works from within SGX enclaves, allowing enclaves to de-randomize the host application.
In \Cref{sec:attack-aslr-js}, we describe that \SBLeak can even be mounted from JavaScript to break ASLR of the browser.

\subsection{Breaking KASLR}\label{sec:attack-aslr-kernel}
In this section, we show that \SBLeak can reliably break KASLR. 
We evaluate the performance of \SBLeak in three different KASLR breaking attacks.
First, we de-randomize the kernel base address.
Second, we de-randomize the direct-physical map.
Third, we find and classify modules based on detected size.

\paragrabf{De-randomizing the Kernel Base Address.}
Jang~\etal\cite{Jang2016} state that the kernel text segment is mapped at a \SI{16}{\mega\byte} boundary somewhere in the \texttt{0xffffffff80000000} - \texttt{0xffffffffc0000000} range.
Given that range, the maximum kernel size is \SI{1}{\giga\byte}.
Combined with the \SI{16}{\mega\byte} alignment, the kernel can only be mapped at one of 64 possible offsets, \ie 6 bits of entropy.
This contradicts the official documentation in The Linux Kernel Archive~\cite{KernelMemoryMap}, which states that the kernel text segment is mapped somewhere in the \texttt{0xffffffff80000000 - 0xffffffff9fffffff} range, giving us a maximum size of \SI{512}{\mega\byte}.
Our experiments show that Jang~\etal\cite{Jang2016} is correct with the address range, but that the kernel is aligned at a 8 times finer \SI{2}{\mega\byte} boundary.
We verified this by checking \texttt{/proc/kallsyms} after multiple reboots.
With a kernel base address range of \SI{1}{\giga\byte} and a \SI{2}{\mega\byte} alignment, we get 9 bits of entropy, allowing the kernel to be placed at one of 512 possible offsets.

Using \SBLeak, we now start at the lower end of the address range and test all of the 512 possible offsets.
If the kernel is mapped at a tested location, we will observe a cache hit.
In our experiments, we see the first cache hit at exactly the same address given by \texttt{/proc/kallsyms}.
Additionally, we see cache hits on all \SI{2}{\mega\byte} aligned pages that follow.
This indicates a \SI{1}{\mega\byte} aliasing effect, supporting the claim made by Islam~\etal\cite{Islam2019Spoiler}.
We only observe the hit on \SI{2}{\mega\byte} aligned pages that follow, as this is our step size.

\cref{tab:kaslr_break} shows the performance of \SBLeak in de-randomizing kernel ASLR.
We evaluated our attack on both an Intel Skylake i7-6600U (without KPTI) and a new Intel Coffee Lake i9-9900K that already includes fixes for Meltdown~\cite{Lipp2018meltdown} and Foreshadow~\cite{Vanbulck2018foreshadow}.
We evaluated our attack on both Windows and Linux, achieving similar results although the ranges differ on Windows.
On Windows, the kernel also starts at a \SI{2}{\mega\byte} boundary, but the possible range is \texttt{0xfffff80000000000} - \texttt{0xfffff80400000000}, which leads to 8192 possible offsets, \ie 13 bits of entropy~\cite{Jang2016}. 

For the evaluation, we tested 10 different randomizations (\ie 10 reboots), each one 100 times, giving us \SIx{1000} samples.
For evaluating the effectiveness of our attack, we use the F1-score.
On the i7-6600U and the i9-9900K, the F1-score for finding the kernel ASLR offset is \SIx{1} when testing every offset a single time, indicating that we always find the correct offset.
In terms of performance, we outperform the previous state of the art~\cite{Jang2016} even though we have an 8 times larger search space.
Furthermore, to evaluate the performance on a larger scale, we tested a single offset \SIx{100} million times.
In that test, the F1-score was 0.9996, showing that \SBLeak virtually always works.
The few misses that we observe are possibly due to the store buffer being drained or that our test program was interrupted.

\begin{table}[t]
  \setlength{\aboverulesep}{0pt}
  \setlength{\belowrulesep}{0pt}
    \caption{Evaluation of \SBLeak in finding the kernel base address and direct-physical map, and kernel modules.
    Number of retries refers to the maximum number of times an offset is tested and number of offsets denotes the maximum number of offsets that need to be tried.}\label{tab:kaslr_break}
\begin{center}
\vspace{-0.175cm}
\adjustbox{max width=\hsize}{{
    \setlength\tabcolsep{1.5pt}
    \begin{tabular}{r ccccc}
      \diagbox{\textbf{Processor}}{\textbf{Target}} &\,\, & \textbf{\#Retries} & \textbf{\#Offsets} & \textbf{Time} & \textbf{F1-Score} \\
      \toprule \vspace{-0.3cm}\\
      \multirow{3}{*}{Skylake (i7-6600U)} & base & 1 & 512 & \SI{72}{\micro\second} & \SIx{1}  \\
                                          \cdashline{2-6}
                                          & direct-physical & 3 & 64000 & \SI{13.648}{\milli\second} & \SIx{1} \\
                                          \cdashline{2-6}
                                          & module & 32 & 262144 & \SI{1.713}{\second} & \SIx{0.98} \\
      \midrule
      \multirow{3}{*}{Coffee Lake (i9-9900K)} & base & 1 & 512 & \SI{42}{\micro\second} & \SIx{1} \\
                                          \cdashline{2-6}
                                          & direct-physical & 3 & 64000 & \SI{8.61}{\milli\second} & \SIx{1} \\
                                          \cdashline{2-6}
                                          & module & 32 & 262144 & \SI{1.33}{\second} & \SIx{0.96} \\
      \bottomrule
    \end{tabular}
  }}
  \end{center}
\vspace{-0.3cm}
\end{table}

\paragrabf{De-randomizing the Direct-physical Map.}
In \cref{sec:background:aslr}, we discussed that the Linux kernel has a direct-physical map that maps the entire physical memory into the kernel virtual address space.
To impede attacks that require knowledge about the map placement in memory, the map is placed at a random offset within a given range at every boot.
According to The Linux Kernel Archive~\cite{KernelMemoryMap}, the address range reserved for the map is \texttt{0xffff888000000000- 0xffffc87fffffffff}, \ie a \SI{64}{\tera\byte} region.
The Linux kernel source code indicates that the map is aligned to a \SI{1}{\giga\byte} boundary.
This gives us $2^{16}$ possible locations.

Using this information, we now use \SBLeak to recover the location of the direct-physical map.
\cref{tab:kaslr_break} shows the performance of the recovery process.
For the evaluation, we tested 10 different randomizations of the kernel (\ie 10 reboots) and for each offset, we repeated the detection 100 times. 
On the Skylake i7-6600U, we were able to recover the offset in under \SI{14}{\milli\second} if KPTI is disabled.
On the Coffee Lake i9-9900K, where KPTI is no longer needed, we were able to do it in under \SI{9}{\milli\second}.

\paragrabf{Finding and Classifying Kernel Modules.}
The kernel reserves \SI{1}{\giga\byte} for modules and loads them at \SI{4}{\kilo\byte}-aligned offset.
In a first step, we can use \SBLeak to detect the location of modules by iterating over the search space in \SI{4}{\kilo\byte} steps.
As kernel code is always present and modules are separated by unmapped addresses, we can detect where a module starts and ends.
In a second step, we use this information to estimate the size of all loaded kernel modules.
The world-readable \emph{/proc/modules} file contains information on modules, including name, size, number of loaded instances, dependencies on other modules, and load state.
For privileged users, it additionally provides the address of the module.
We correlate the size from \emph{/proc/modules} with the data from our \SBLeak attack and can identify all modules with a unique size.
On the i7-6600U, running Ubuntu 18.04 with kernel version 4.15.0-47, we have a total of 26 modules with a unique size.
On the i9-9900K, running Ubuntu 18.10 with kernel version 4.18.0-17, we have a total of 12 modules with a unique size.
\cref{tab:kaslr_break} shows the accuracy and performance of \SBLeak for finding and classifying those modules.

\paragrabf{The Strange Case of Non-Canonical Addresses.}
For valid kernel addresses, there are no false positives with \SBLeak. 
Interestingly, when used on a non-canonical address, \ie an address where the bits 47 to 63 are not all `0' or `1', \SBLeak reports this address to be backed by a physical page. 
However, these addresses are invalid by definition and can thus never refer to a physical address~\cite{Intel_vol3}. 
We guess that there might be a missing check in the store-to-load forwarding unit, which allows non-canonical addresses to enter the store buffer and be forwarded to subsequent loads despite not having a physical address associated. 
Although the possibility of such a behaviour is documented~\cite{HilyZH09}, it is still unexpected, and future work should investigate whether it could lead to security problems.

\paragrabf{Comparison to Other Side-Channel Attacks on KASLR.} 
Previous microarchitectural attacks on ASLR relied on address-translation caches~\cite{Hund2013,Jang2016,Gruss2016Prefetch,Gras2018} or branch-predictor states~\cite{Evtyushkin2016ASLR,Evtyushkin2018}. 
We compare \SBLeak against previous attacks on kernel ASLR~\cite{Hund2013,Jang2016,Gruss2016Prefetch,Evtyushkin2016ASLR}. 

\begin{table}
  \caption{Comparison of microarchitectural attacks on KASLR. Of all known attacks, \SBLeak is by far the fastest and in contrast to all other attacks has no requirements.}
  \label{tab:kaslr-compare}
\adjustbox{max width=\hsize}{
 \begin{tabular}{l|rcl}
  \textbf{Attack} & \textbf{Time} & \textbf{Accuracy} & \textbf{Requirements} \\ \midrule
  Hund~\etal\cite{Hund2013} & \SI{17}{\second} & \SI{96}{\percent} & - \\
  Gruss~\etal\cite{Gruss2016Prefetch} & \SI{500}{\second} & N/A & cache eviction \\
  Jang~\etal\cite{Jang2016} & \SI{5}{\milli\second} & \SI{100}{\percent} & Intel TSX \\
  Evtyushkin~\etal\cite{Evtyushkin2016ASLR} & \SI{60}{\milli\second} & N/A & BTB reverse engineering \\
   \SBLeak (our attack) & \SI{42}{\textbf{\micro\second}} & \SI{100}{\percent} & - \\
 \end{tabular}
 }
\end{table}

\Cref{tab:kaslr-compare} shows that our attack is the fastest attack that works on any Intel x86 CPU.
In terms of speed and accuracy, \SBLeak is similar to the methods proposed by Jang~\etal\cite{Jang2016} and Evtyushkin~\etal\cite{Evtyushkin2016ASLR}. 
However, one advantage of \SBLeak is that it does not rely on CPU extensions, such as Intel TSX, or precise knowledge of internal data structures, such as the reverse-engineering of the branch-target buffer (BTB). 
In particular, the attack by Jang~\etal\cite{Jang2016} is only applicable to selected CPUs starting from the Haswell microarchitecture, \ie it cannot be used on any CPU produced earlier than 2013. 
Similarly, Evtyushkin~\etal\cite{Evtyushkin2016ASLR} require knowledge of the internal workings of the branch-target buffer, which is not known for any microarchitecture newer than Haswell. 
\SBLeak works regardless of the microarchitecture, which we have verified by successfully running it on microarchitectures starting from Pentium 4 Prescott (released 2004) to Whiskey Lake and Coffee Lake R (both released end of 2018) (\cf \Cref{tab:setups}).

\subsection{Inferring ASLR from SGX}\label{sec:attack-aslr-sgx}
\SBLeak does not only work for privileged addresses (\cf \Cref{sec:attack-aslr-kernel}), it also works for otherwise inaccessible addresses, such as SGX enclaves. 
We demonstrate 3 different scenarios of how \SBLeak can be used in combination with SGX. 

\paragrabf{\textbf{\SBLeak from Host to Enclave.}}
With \SBLeak, it is possible to detect enclaves in the EPC (Enclave Page Cache), the encrypted and inaccessible region of physical memory reserved for Intel SGX.
Consequently, we can de-randomize the virtual addresses used by SGX enclaves. 

However, this scenario is artificial, as an application has more straightforward means to determine the mapped pages of its enclave. 
First, applications can simply check their virtual address space mapping, \eg using \texttt{/proc/self/maps} in Linux. 
Second, reading from EPC pages always returns `-1'~\cite{sgxdeveloperref}, thus if a virtual address is also not part of the application, it is likely that it belongs to an enclave.
Thus, an application can simply use TSX or install a signal handler to probe its address space for regions which return `-1' to detect enclaves. 

For completeness, we also evaluated this scenario. 
In our experiments, we successfully detected all pages mapped by an SGX enclave in our application using \SBLeak. 
Like with KASLR before, we had no false positives, and the accuracy was \SI{100}{\percent}. 

\paragrabf{\SBLeak from Enclave to Host.}
While the host application cannot access memory of an SGX enclave, the enclave has full access to the virtual memory of its host. 
Schwarz~\etal\cite{Schwarz2019SGXMalware} showed that this asymmetry allows building enclave malware by overwriting the host application's stack for mounting a return-oriented programming exploit. 
One of the primitives they require is a possibility to scan the address space for mapped pages without crashing the enclave. 
While this is trivial for normal applications using syscalls (\eg by abusing the \texttt{access} syscall, or by registering a user-space signal handler), enclaves cannot rely on such techniques as they cannot execute syscalls. 
Thus, Schwarz~\etal propose \emph{TAP}, a primitive relying on TSX to test whether an address is valid without the risk of crashing the enclave. 

The same behavior can also be achieved using \SBLeak without having to rely on TSX. 
By leveraging speculative execution to induce transient execution, not a single syscall is required to mount \SBLeak. 
As the \texttt{rdtsc} instruction cannot be used inside enclaves, we use the same counting-thread technique as Schwarz~\etal\cite{Schwarz2017SGX}. 
The resolution of the counting thread is high enough to mount a \FlushReload attack inside the enclave. 
In our experiments, a simple counting thread achieved \SIx{0.92} increments per cycle on a Skylake i7-8650U, which is sufficient to mount a reliable attack. 

With \SI{290}{MB/s}, the speed of \SBLeak in an enclave is a bit slower than the speed of \emph{TAP}~\cite{Schwarz2019SGXMalware}. 
The reason is that \SBLeak requires a timer to mount a \FlushReload attack, whereas \emph{TAP} does not require any timing or other side-channel information. 
Still, \SBLeak is a viable alternative to the TSX-based approach for detecting pages of the host application from within an SGX enclave, in particular as many processors do not support TSX.

\paragrabf{\SBLeak from Enclave to Enclave.}
Enclaves are not only isolated from the operating system and host application, but enclaves are also isolated from each other. 
Thus, \SBLeak can be used from a (malicious) enclave to infer the address-space layout of a different, inaccessible enclave. 

We evaluated the cross-enclave ASLR break using two enclaves, one benign and one malicious enclave, started in the same application. 
Again, to get accurate timestamps for \FlushReload, we used a counting thread. 

As the enclave-to-enclave scenario uses the same code as the enclave-to-host scenario, we also achieve the same performance.

\subsection{Breaking ASLR from JavaScript}\label{sec:attack-aslr-js}
\SBLeak cannot only be used from unprivileged native applications but also in JavaScript to break ASLR in the browser. 
In this section, we evaluate the performance of \SBLeak from JavaScript running in a modern browser. 
Our evaluation was done on Google Chrome 70.0.3538.67 (64-bit) and Mozilla Firefox 66.0.2 (64-bit). 

There are two main challenges for mounting \SBLeak from JavaScript. 
First, there is no high-resolution timer available. 
Therefore, we need to build our own timing primitive. 
Second, as there is no flush instruction in JavaScript, \FlushReload is not possible. 
Thus, we have to resort to a different covert channel for bringing the microarchitectural state to the architectural state. 

\paragrabf{Timing Primitive.}
To measure timing with a high resolution, we rely on the well-known use of a counting thread in combination with shared memory~\cite{Schwarz2017Timers,Gras2017}.
As Google Chrome has re-enabled \texttt{SharedArrayBuffers} in version 67\footnote{\url{https://bugs.chromium.org/p/chromium/issues/detail?id=821270}}, we can use the existing implementations of such a counting thread. 

In Google Chrome, we can also use \texttt{BigUint64Array} to ensure that the counting thread does not overflow. 
This improves the measurements compared to the \texttt{Uint32Array} used in previous work~\cite{Schwarz2017Timers,Gras2017} as the timestamp is increasing strictly monotonically. 
In our experiments, we achieve a resolution of \SI{50}{\nano\second} in Google Chrome, which is sufficient to distinguish a cache hit from a miss. 

\paragrabf{Covert Channel.}
As JavaScript does not provide a method to flush an address from the cache, we have to resort to eviction as shown in previous work~\cite{Oren2015,Schwarz2017Timers,Gras2017,Vila2018eviction,Kocher2019}. 
Thus, our covert channel from the microarchitectural to the architectural domain, \ie the decoding of the leaked value which is encoded into the cache, uses \EvictReload instead of \FlushReload. 

For the sake of simplicity, we can also just access an array which has a size 2-3 times larger than the last-level cache to ensure that the array is evicted from the cache. 
For our proof-of-concept, we use this simple approach as it is robust and works for the attack. 
While the performance increases significantly when using targeted eviction, we would require 256 eviction sets.
Building them would be time consuming and prone to errors. 

\paragrabf{Illegal Access.}
In JavaScript, we cannot access an inaccessible address architecturally. 
However, as JavaScript is compiled to native code using a just-in-time compiler in all modern browsers, we can leverage speculative execution to prevent the fault. 
Hence, we rely on the same code as Kocher~\etal\cite{Kocher2019} to speculatively access an out-of-bounds index of an array. 
This allows to iterate over the memory (relative from our array) and detect which pages are mapped and which pages are not mapped. 

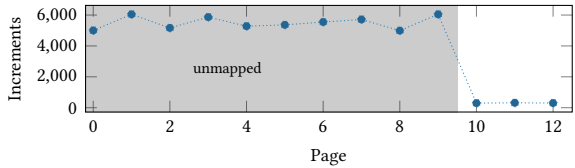
\begin{figure}
 \input{img/js_timings.tikz}
 \caption{\SBLeak with \EvictReload in JavaScript clearly shows whether an address (relative to a base address) is backed by a physical page and thus valid.}
 \label{fig:js-timing}
\end{figure}

\paragrabf{Full Exploit.}
When putting everything together, we can distinguish for every location relative to the start array whether it is backed by a physical page or not.
Due to the limitations of the JavaScript sandbox, especially due to the slow cache eviction, the speed is orders of magnitude slower than the native implementation, as it can be seen in \Cref{fig:js-timing}.
Still, we can detect whether a virtual address is backed by a physical page within \SI{450}{\milli\second}, making \SBLeak also realistic from JavaScript.

\section{\AttackPrimitive}\label{sec:attack-cf} %
While \SBLeak can already be used for powerful attacks, \AttackPrimitive, the TLB-augmented variant of \SBLeak, allows for even more powerful attacks as we show in this section. 
So far, most microarchitectural attacks which can attack the kernel, such as \PrimeProbe~\cite{Schwarz2018KeyDrown} or DRAMA~\cite{Pessl2016}, require at least some knowledge of physical addresses. 
Since physical address information is not provided to unprivileged application, these attacks either require additional side channels~\cite{Schwarz2018KeyDrown,Gruss2018Rowhammer} or have to blindly attack targets until the correct target is found~\cite{Schwarz2017SGX}. 

With \AttackPrimitive we can directly retrieve side-channel information for any target virtual address, regardless of whether it can be accessed in the current privilege level or not. 
In particular, we can detect whether a virtual address has a valid translation in either the iTLB or dTLB. 
This allows an attacker to infer whether an address was recently used. 

\AttackPrimitive can attack both the iTLB and dTLB. 
Using \AttackPrimitive, an attacker can detect recently accessed \textit{data pages} on the current hyperthread. 
Moreover, an attacker can also detect \textit{code pages} recently used for instruction execution on the current hyperthread. 

As the measurement with \AttackPrimitive results in a valid mapping of the target address, we also require a method to evict the TLB. 
While this can be as simple as accessing (dTLB) or executing (iTLB) data on more pages than there are entries in the TLB, this is not an optimal strategy. 
Instead, we rely on the reverse-engineered eviction strategies from Gras~\etal\cite{Gras2018}. 

The attack process is the following:
\begin{itemize}
 \item[\circleds{1}] Build eviction sets for the target address(es)
 \item[\circleds{2}] \AttackPrimitive on the target address(es) to detect activity
 \item[\circleds{3}] Evict address(es) from iTLB and dTLB
 \item[\circleds{4}] Goto 2
\end{itemize}

In \Cref{sec:attack-cf-tsx}, we show that \AttackPrimitive allows an unprivileged application to spy on TSX transactions. 
In \Cref{sec:attack-cf-kernel}, we demonstrate an attack on the Linux kernel using \AttackPrimitive.

\subsection{Breaking the TSX Atomicity}\label{sec:attack-cf-tsx}
Intel TSX guarantees the atomicity of all instructions and data accesses which are executed inside a TSX transaction. 
Thus, Intel TSX has been proposed as a security mechanism against side-channel attacks~\cite{Guan2015,Gruss2017TSX}. 

With \AttackPrimitive, an attacker can break the atomicity of TSX transactions by recovering the TLB state after the transaction. 
While Intel TSX reverts the effect of all instructions, including the invalidation of modified cache lines, the TLB is not affected. 
Thus, \AttackPrimitive can be used to detect which addresses are valid in the TLB, and thus infer which addresses have been accessed within a transaction. 
We verified that this is not only possible after a successful commit of the transaction, but also after a transaction aborted. 

As a consequence, an attacker can abort a transaction at an arbitrary point (\eg by causing an interrupt or a conflict in the cache) and use \AttackPrimitive to detect which pages have been accessed until this point in time. 
From that, an attacker can learn memory access patterns, which should be invisible due to the guarantees provided by Intel TSX.

\subsection{Inferring Control Flow of the Kernel}\label{sec:attack-cf-kernel}
The kernel is a valuable target for attackers, as it processes all user inputs coming from I/O devices. 
Microarchitectural attacks targeting user input directly in the kernel usually rely on \PrimeProbe~\cite{Ristenpart2009,Oren2015,Schwarz2018KeyDrown,Monaco2018} and thus require knowledge of physical addresses.

With \AttackPrimitive, we do not require knowledge of physical addresses to spy on the kernel. 
In the following, we show that \AttackPrimitive can spy on any type of kernel activity.
We illustrate this with the examples of mouse input and Bluetooth events.

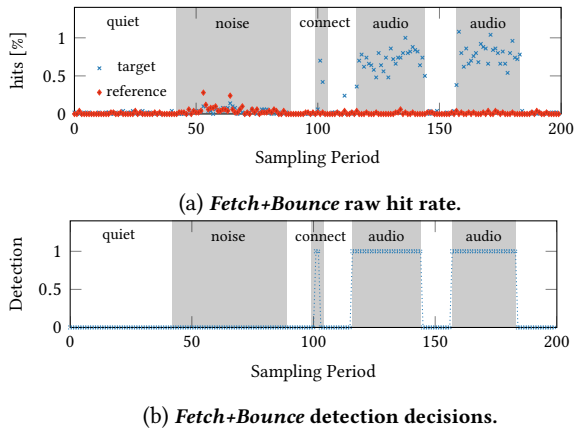
\begin{figure}
 \begin{subfigure}{\linewidth}
  \input{img/bt_access_trace_hits.tikz}
  \caption{\AttackPrimitive raw hit rate.}
  \label{fig:bluetooth-events-hits}
 \end{subfigure}
 \begin{subfigure}{\linewidth}
  \input{img/bt_access_trace_detection.tikz}
  \caption{\AttackPrimitive detection decisions.}
  \label{fig:bluetooth-events-det}
 \end{subfigure}
	
 \caption{Detecting Bluetooth events by monitoring TLB hits via \AttackPrimitive on pages at the start of the \texttt{bluetooth} kernel module.}
 \label{fig:bluetooth-events}
\end{figure}

As a simple proof of concept we monitor the first 8 pages of a target kernel module.
To obtain a baseline for the general kernel activity, and thus the TLB activity for kernel pages, we also monitor one reference page from a kernel module that is rarely used (in our case \texttt{i2c\_i801}).
By comparing the activity on the 8 pages of the kernel module to the baseline, we can determine whether the kernel module is currently actively used or not.
To get the best results we use \AttackPrimitive with both the iTLB and dTLB. 
This makes the attack independent of the type of activity in the kernel module, \ie there is no difference if code is executed or data is accessed.
Our spy changes its hyperthread after each \AttackPrimitive measurement.
This reduces the attack's resolution, but allows it to detect activity on all hyperthreads.
For further processing, we sum the resulting TLB hits over a \textit{sampling period} which consists of \SIx{5000} measurements, and then apply a basic detection filter to this sum:
We calculate the ratio of hits on the target pages and the reference page.
If the number of hits on the target pages is above a sanity lower bound and more importantly, above the number of cache hits on the reference page, \ie above the baseline, then, the page was recently used (\cf \Cref{fig:bluetooth-events-det}).

\paragrabf{Detecting User Input.}
We now investigate how well \AttackPrimitive works for spying on input-handling code in the kernel. 
While Schwarz~\etal\cite{Schwarz2018KeyDrown} attacked the kernel code for PS/2 keyboards and laptops, we target the kernel module for USB human-interface devices. 
This has the advantage that we can attack input from a large variety of modern USB input devices. 

We first locate the kernel module using \SBLeak as described in \Cref{sec:attack-aslr-kernel}.
With 12 pages (kernel 4.15.0), the kernel module does not have a unique size among all modules but is one of only 3. 
Thus, we can either try to identify the correct module or monitor all of them. 

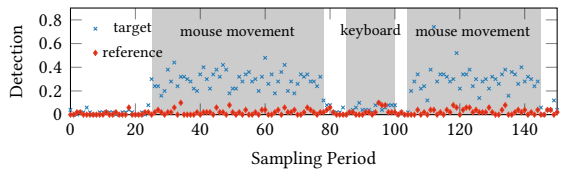
\begin{figure}
 \input{img/usbhid_access_trace.tikz}
 \caption{Mouse movement detection. The mouse movements are clearly detected. The USB keyboard activity does not cause more TLB hits than observed as a baseline.}
 \label{fig:mouse-events}
\end{figure}

\Cref{fig:mouse-events} shows the results of using \AttackPrimitive on a page of the \texttt{usbhid} kernel module. 
It can be clearly seen that mouse movement results in a higher number of TLB hits.
USB keyboard input, however, seems to fall below the detection threshold with our simple method.
Given this attack's low temporal resolution, repeated accesses to a page are necessary for clear detection. 
Previous work has shown that such an event trace can be used to infer user input, \eg URLs~\cite{Oren2015,Lipp2017Interrupt}. 

\paragrabf{Detecting Bluetooth Events.}
Bluetooth events can give valuable information about the user's presence at the computer, \eg connecting (or disconnecting) a device usually requires some form of user interaction.
Tools, such as Dynamic Lock on Windows 10~\cite{Win10DynamicLock}, use connect and disconnect events to unlock and lock a computer automatically. 
Thus, these events are apparently a useful indicator to detect whether the user is currently using the computer. 
Also, it is useful to monitor these events as a trigger signal for UI redressing attacks.

To spy on these events, we first locate the Bluetooth kernel module using \SBLeak. 
As the Bluetooth module is rather large (134 pages on kernel 4.15.0) and has a unique size, it is easy to distinguish it from other kernel modules. 

\Cref{fig:bluetooth-events} shows a \AttackPrimitive trace while generating Bluetooth events. 
While there is a constant noise floor due to TLB collisions (\Cref{fig:bluetooth-events-hits}), we can see a clear increase in TLB hits on the target address for every Bluetooth event.
After applying our detection filter, we can detect events such as connecting and playing audio over the Bluetooth connection with a high accuracy (\Cref{fig:bluetooth-events-det}).

Our results indicate that the precision of the detection and distinction of events with \AttackPrimitive can be significantly improved.
Future work should investigate profiling code pages of kernel modules, similar to previous template attacks~\cite{Gruss2015Template}.

\section{Leaking Kernel Memory}\label{sec:attack-leak}%
In this section, we present \DataLeak, a novel covert channel to leak memory using Spectre.
Most Spectre attacks, including the original Spectre attack, use the cache as a covert channel to encode values leaked from the kernel~\cite{Kocher2019,Maisuradze2018spectre5,Kiriansky2018speculative,Koruyeh2018spectre5,Horn2018spectre4,OKeeffe18sgxspectre,Chen2018SGXpectre,Schwarz2019netspectre}. 
Other covert channels for Spectre attacks, such as port contention~\cite{Bhattacharyya2019} or AVX~\cite{Schwarz2019netspectre} have since been presented.
However, it is unclear how commonly such gadgets can be found and can be exploited in real-world software. 

With \DataLeak, we show how the TLB effects on the store buffer (\cf \Cref{sec:attack-cf}) can be combined with speculative execution to leak kernel data.
We show that any cache-based Spectre gadget can be used for \DataLeak.
As the secret-dependent page access also populates the TLB, such a gadget also encodes the information in the TLB.
With \SBLeak, we can then reconstruct which of the pages was accessed and thus infer the secret.

\begin{listing}[t]
 \begin{lstlisting}
if ( index < bounds )
    y = oracle[ data[index] * 4096 ]; \end{lstlisting}
 \caption{A simple Spectre-PHT gadget, which allows speculative access of \texttt{data} out of bounds and encodes the value in \texttt{oracle}.}
 \label{lst:spectre-gadget}

\end{listing}

While at first, the improvements over the original Spectre attack might not be obvious, there are 2 huge advantages. 

\paragrabf{Advantage 1: It requires less control over the Spectre gadget.}
First, for \DataLeak, an attacker requires less control over the Spectre gadget. 
In a Spectre-PHT (aka Spectre Variant 1) attack, a gadget similar to the one illustrated in \Cref{lst:spectre-gadget} is required. 
There, an attacker requires full control over \texttt{index}, and also certain control over \texttt{oracle}. 
Specifically, the base address of \texttt{oracle} has to point to user-accessible memory which is shared between attacker and victim. 
Furthermore, the base address has to either be known or be controlled by the attacker. 
This limitation potentially reduces the number of exploitable gadgets. 

\paragrabf{Advantage 2: It requires no shared memory.}
Second, with \DataLeak, we get rid of the shared-memory requirement. 
Especially on modern operating systems, shared memory is a limitation, as these operating systems provide stronger kernel isolation~\cite{Gruss2017KASLR}. 
On such systems, only a few pages are mapped both in user and kernel space, and they are typically inaccessible from the user space. 
Moreover, the kernel can typically not access user space memory due to supervisor mode access prevention (SMAP). 
Hence, realistic Spectre attacks have to resort to \PrimeProbe~\cite{Trippel2018MeltdownPrime}. 
However, \PrimeProbe requires knowledge of physical addresses, which is not exposed on modern operating systems. 

With \DataLeak, it is not necessary to have a memory region which is user accessible and shared between user and kernel space. 
For \DataLeak, it is sufficient that the base address of \texttt{oracle} points to a kernel address which is also mapped in user space.
Even in the case of KPTI~\cite{LWN_kpti}, there are still kernel pages mapped in the user space.
On kernel 4.15.0, we identified 65536 such kernel pages (\ie \SI{256}{\mega\byte}) when KPTI is enabled, and multiple gigabytes when KPTI is disabled.
Hence, \texttt{oracle} only has to point to any such range of mapped pages.
Thus, we expect that there are simpler Specter gadgets which are sufficient to mount this attack.

\subsection{Leaking Data}

To evaluate \DataLeak, we use a custom \texttt{ioctl} in the Linux kernel containing a Spectre gadget as illustrated in \Cref{lst:spectre-gadget}.
The \texttt{oracle} array points to a kernel address which is also mapped in the user space but not user accessible.
Furthermore, we can control \texttt{index} from user space, \ie it is provided as an argument to the \texttt{ioctl}.

By first providing in-bounds values for \texttt{index}, we mistrain the branch predictor in the kernel in-place~\cite{Canella2019A}.
Then, by providing an out-of-bounds value for \texttt{index}, the gadget encodes the speculatively accessed value in the TLB.
Finally, in  user space, we use \SBLeak to detect which kernel page of \texttt{oracle} has a valid TLB entry.
The TLB entry directly depends on the secret leaked from the kernel.

We cannot ensure that speculative execution always misspeculates, hence, we have to re-run \DataLeak multiple times per byte to leak. 
As with \AttackPrimitive (\cf \Cref{sec:attack-cf}), this again requires TLB eviction after every run of \DataLeak. 
With this approach, we can reliably leak data from the kernel in the same way and with the same efficiency as in the original Spectre attack~\cite{Kocher2019}.
However, we reduced the requirements for an attacker significantly, as our approach also works with active SMAP and there is no need for shared memory. 

\section{Microarchitectural Root-Cause Analysis}\label{sec:analysis}%
We now peform a microarchitectural root-cause analysis to distinguish \SBLeak from the \emph{write transient forward} effect described in the Fallout paper~\cite{Canella2019Fallout}.
Generally, store-to-load forwarding uses two components: the store buffer and the memory disambiguation predictor.
As store-to-load forwarding intends to forward data from a previous store that has not yet been written to the memory subsystem to a subsequent load from the same address, we can distinguish four different cases.

The first case is a \textbf{true positive match}.
In the true positive match case, the store buffer contains an entry that potentially matches with the full physical address.
The store-to-load forwarding logic hence forwards the data to the load as this is expected to be the correct behavior.

The second case is then the \textbf{true negative match.}
Contrary to the true positive match, the store buffer does not find a potentially matching entry and, indeed, there is no match with a full physical address.
The store buffer then does not forward any information as this is again the expected behavior.
In this work, we exploit this as negative information in combination with the true positive case to determine the validity of addresses.

The third is the \textbf{false negative match} case.
In this case, the store buffer finds no matching store even though a full physical address match existed.
As expected, no forwarding takes place and the executed load works on outdated values, \eg from the L1 cache.
One likely situation for this was shown by Horn~\cite{Horn2018spectre4} where the load operation was scheduled earlier than the store it depends on, hence no store buffer entry can exist and the load transiently uses stale data.

The final case is then the \textbf{false positive match}.
At first, the store buffer finds a matching store, which later on turns out to not be a full physical address match.
Nevertheless, the load continues as a zombie load before being squashed and transiently forwards data to dependent instructions.
Islam~\etal\cite{Islam2019Spoiler} exploited this behavior to obtain physical-address information via a timing attack.
In its essence, this is the \emph{write transient forwarding effect} exploited in the Fallout attack~\cite{Canella2019Fallout}.

Based on this analysis, we can conclude that the effects used in this paper clearly differs from the effect exploited in Fallout with the first exploiting a combination of the true positive and negative match and the second one exploiting the false positive match.
Note that a similar analysis has been performed by Canella~\etal\cite{Canella2020attacks}.

\paragraph{Experimental Verification}
Designing an experiment on modern CPUs to verify experimentally that the effects have different root causes is difficult. 
However, for this evaluation, we can rely on older microarchitecture. 
We used a Pentium 4 531 from 2004 running Ubuntu 13.04 (kernel 3.8.0) with all mitigations disabled. 
The Pentium 4 already supports out-of-order execution and speculative execution. 
We verified that our CPU is vulnerable to transient-execution attacks by successfully running a Spectre-PHT and Spectre-BTB PoC~\cite{Canella2019A}.

On the Pentium 4, we can reliably reproduce \SBLeak, \ie the cases of a true positive and true negative match. 
This result shows that Meltdown-type effects are already observable on the pre-Core microarchitectures. 
However, the CPU is not vulnerable to Fallout, \ie there is no false positive match in the store buffer. 
Hence, from this experiment we can conclude that the effects observed in this paper and in the Fallout paper are indeed different effects, and that the \emph{write transient forwarding effect} exploited in Fallout was introduced only with later microarchitectures. 

\paragraph{Offset Match}
Fallout~\cite{Canella2019Fallout} describes the \textit{write transient forward} effect to forward the previous store transiently to the load with the exact \textit{same} page offset.
However, this is not entirely accurate.
In our experiments, we have observed this effect as long as the page offset of the load is within the range (\textit{store\_offset}, \textit{store\_offset+store\_size}) where \textit{store\_offset} is the page offset of the store instruction and \textit{store\_size} the number of bytes stored (\eg 1 byte for a \texttt{movb} store).
This shows that not only the page offset is taken into account but also the number of bytes actually written by the store instruction.
For example, if the store operation is a \texttt{movq} instruction the page offset of the faulting load operation can be up to 7 bytes higher.
Further, we experimentally verified that we can index a store to an \texttt{XMM} register (128 bit) with a different offset up to 15 and to an \texttt{YMM} register (256 bit) up to 31 bytes.

\section{Discussion \& Related Work}\label{sec:discussion}%

With \SBLeak, we demonstrate a powerful side-channel attack to detect whether a virtual address has a valid mapping by exploiting store-to-load forwarding. 
While similar attacks are known, \SBLeak is both the most reliable and fastest attack so far. 
The attack which comes closest in terms of reliability and performance is the TSX-based attack by Jang~\etal\cite{Jang2016}. 
However, TSX is only supported by 34\% of the Core CPUs and 43\% of the Xeon CPUs released since 2013, which are currently available for sale~\cite{GeizhalsTSX2019}.
\SBLeak does work both with and without TSX. 
When leveraging TSX for \SBLeak, the performance increases by factor 4 which even outperforms Jang~\etal\cite{Jang2016}. 
The F1-score, \ie both precision and recall, are not affected regardless of whether TSX is used or not. 
As a result, \SBLeak is applicable to a much wider range of CPUs, and also in restricted environments such as JavaScript. 
Thus, \SBLeak can also be used in similar scenarios as the JavaScript-based ASLR break by Gras~\etal\cite{Gras2017}. 

Gras~\etal\cite{Gras2018} demonstrated that sharing the TLB across hyperthreads enables eviction-based attacks on the TLB. 
However, this attack requires to first reverse engineer the TLB for building precise eviction sets of the target mapping. 
\AttackPrimitive is a similar side-channel attack but does not require knowledge of TLB sets.  
\AttackPrimitive works directly with the virtual address of the target mapping, thus reducing the noise of addresses mapping to the same set in the TLB. 
On a high level, Gras~\etal\cite{Gras2018} showed a \PrimeProbe-type attack on the TLB, whereas we show a more precise \EvictReload-type attack on the TLB. 

With \DataLeak, we show a fast and practical covert channel for Spectre attacks which can replace \FlushReload in certain scenarios. 
While other covert channels have been proposed~\cite{Bhattacharyya2019,Kiriansky2018speculative,Schwarz2019netspectre,Trippel2018MeltdownPrime}, \FlushReload is still the most reliable covert channel.
As \DataLeak can exploit the same gadgets, and gadgets with fewer requirements, we already know from previous work that such real-world gadgets exist and have been exploitable in the wild~\cite{Kocher2019}. 
In line with previous Spectre papers, finding new gadgets is an orthogonal problem and thus is not discussed in this paper. 

Although the information leakage is caused by the hardware, countermeasures against \SBLeak are possible. 
The basic idea is to not have different security domains in the same address space. 
Since Meltdown, KAISER~\cite{Gruss2017KASLR} was deployed on all modern operating systems to ensure that the kernel is not mounted in the user's address space. 
This prevents Meltdown but also claims to prevent side-channel attacks on kernel addresses~\cite{Hund2013,Gruss2016Prefetch,Jang2016} including \SBLeak, \AttackPrimitive, and \DataLeak. 
As Canella~\etal\cite{Canella2020kaslr} have shown does this claim not hold completely as they were still able to de-randomize the kernel.
They analyzed the commonalities of side-channel attacks on kernel addresses and propose FLARE to mitigate such attacks.
Unfortunately, Koschel\etal~\cite{koschel2020} have shown that this is also insufficient.
Additionally, we propose to apply the same principle to SGX and sandboxes, similar to site isolation~\cite{Chromium2018SiteIsolation}. 
In general, future hardware and software designs should ensure that different security domains are not shared in the same address space to reduce the attack surface. 

While preventing the attacks is possible, detecting the attacks is significantly harder. 
Depending on the implementation, \SBLeak does not require any operating-system interaction at all. 
For example, when implemented using speculative execution for exception prevention, no syscall is required, and architecturally, no exception is triggered. 
Thus, \SBLeak is completely invisible to the operating system. 
Several works suggested relying on performance counters to detect ongoing side-channel attacks~\cite{Chiappetta2015,Herath2015,Gruss2016Flush,Payer2016,Zhang2016CloudRadar,Irazoqui2018mascat}, \eg by detecting an anomaly in cache misses. 
However, it is unclear how such detection mechanisms perform in real-world scenarios. 
Especially for the KASLR break, where the total runtime is only \SI{42}{\micro\second}, it is questionable whether this is detectable among the average system noise.
It is even more questionable whether the system could in time respond to a potential ongoing attack.

\section{Conclusion}\label{sec:conclusion}
In this paper, we demonstrated \SBLeak, a Meltdown-like attack on recent patched CPUs.
We showed that correct store-to-load forwarding via the store buffer introduces potent channels to leak data and meta data.
We showed that we can break KASLR on fully patched machines in \SI{42}{\micro\second}.
We demonstrated using the same side channel to also reveal the address space of Intel SGX enclaves, and break ASLR from JavaScript.
In combination with TLB state changes we were able to break the atomicity of TSX as well as monitor the control flow of the kernel.
Finally, we found that Spectre v1 gadgets can also be exploited using \SBLeak. 
We showed that this allows us to leak arbitrary kernel memory in realistic scenarios.

Our work shows that the hardware fixes for Meltdown in Whiskey Lake and Coffee Lake CPUs are clearly not sufficient.
We conclude that software-based isolation of user and kernel space should remain enabled even on the most recent processor generations. 

\ifAnon

\else
\section*{Acknowledgments}
We thank Moritz Lipp and Vedad Had\v{z}i\'{c} from Graz University of Technology and Julian Stecklina from Cyberus Technology for contributing ideas and experiments.
This work has been supported by the Austrian Research Promotion Agency (FFG) via the project ESPRESSO, which is funded by the province of Styria and the Business Promotion Agencies of Styria and Carinthia.
This project has received funding from the European Research Council (ERC) under the European Union's Horizon 2020 research and innovation programme (grant agreement No 681402).
Additional funding was provided by a generous gift from Intel and ARM.
Any opinions, findings, and conclusions or recommendations expressed in this paper are those of the authors and do not necessarily reflect the views of the funding parties.
\fi

{\footnotesize \bibliographystyle{acm-url}
\bibliography{main}}

\end{document}

%% file: img/store_buffer.tikz
\begin{tikzpicture}[yscale=0.7]

\draw (0,2) rectangle +(2,0.75) node[midway] {Store Data};
\draw (0,0) rectangle +(2,0.75) node[midway] {Load Data};

\draw (3,0) rectangle +(2,1) node[midway] {Load Buffer};
\draw[->,>=stealth,ultra thick,in=180,out=0] (2,0.375) to (3,0.5);
\draw (3,1.75) rectangle +(2,1) node[midway] {Store Buffer};
\draw[->,>=stealth,ultra thick,in=180,out=0] (2,2.375) to (3,2.25);

\draw (7,0.5) rectangle +(2,2) node[midway] {\parbox{2cm}{\centering L1\\Data\\Cache}};

\draw[->,>=stealth,ultra thick,in=180,out=0] (5,2.375) to (7,2);
\draw[->,>=stealth,ultra thick,in=0,out=180] (7,1) to (5,0.375);

\draw[->,>=stealth,densely dotted,thick,in=10,out=-10,looseness=3] (5,2) to node[midway,left] {\scriptsize{Store-to-load forwarding}} (5,0.75);

\draw[densely dashed,black!60] (-0.25,-0.25) rectangle +(2.75,3.75) node[black!60,xshift=-1.375cm,yshift=-0.25cm] {\small Execution Unit};

\draw[densely dashed,black!60] (2.75,-0.25) rectangle +(6.75,3.75) node[black!60,xshift=-1.375cm,yshift=-0.25cm] {\small Memory Subsystem};

\end{tikzpicture}

%% file: img/sgx.tikz
\tikzstyle{process} = [rectangle, minimum width=2cm, minimum height=0.65cm, text centered, draw=black, fill=white]
\tikzstyle{arrow} = [thick,->,>=stealth]
\begin{tikzpicture}[yscale=0.65]
\definecolor{msblue}{HTML}{0000FF}

\draw (0.25, 2.25) rectangle node [yshift=1.85cm] {\large Application} +(9,4.75);

\draw [densely dashed] (5.75, 2.5) rectangle node [anchor=north,yshift=2.1cm,xshift=.7cm] {Trusted enclave} +(3, 4);
\draw [fill=gray!20] (5.5, 4) rectangle node [rotate=90,color=black] {Call Gate} +(0.5,2.5);

\draw [densely dashed] (0.75, 2.5) rectangle node [anchor=north,yshift=2.1cm,xshift=-.7cm] {Untrusted} +(3, 4.25);
\node (create) at (2.25,6) [process] {\small Load Enclave};
\draw [arrow] (2.25,5.68) -- (2.25,4.82);
\node (call) at (2.25,4.5) [process] {\small Call Trusted Fnc.};
\draw [arrow] (2.25,4.18) -- (2.25,3.42);
\node (call) at (2.25,3.1) [process] {\small ... };

\draw [arrow] (3.55,4.65) to node[midway,sloped,yshift=.25cm]{EENTER} (5.5,5.5);
\draw [arrow] (6,5.5) -- (6.5,5.5);
\draw (6.5,3.5) edge[out=180,in=0,arrow] node[midway,sloped,yshift=-0.25cm] {EEXIT} (3.55,4.35);
 
\node (create) at (7.5,5.5) [process] {\small Trusted Code};
\node (create) at (7.5,3.5) [process] {\small Return};
\draw [arrow] (7.5,5.17) -- (7.5,3.82);

\end{tikzpicture}

%% file: img/sb-tlb.tikz
\begin{tikzpicture}
\begin{axis}[
style={font=\footnotesize},
xlabel={Page},
ylabel={Repetitions},
width=0.95\hsize,
xmin=-0.2,
xmax=30.5,
scaled y ticks=false,
height=3cm
]
\draw[fill=black, opacity=0.2,draw=none] (axis cs: -0.5,-2) rectangle (axis cs: 7.5,10);
\draw[fill=yellow, opacity=0.2,draw=none] (axis cs: 16.5,-2) rectangle (axis cs: 17.5,10);

\addplot+[densely dotted,mark=*,mark size=1.5pt] table[x=page,y=repeat,col sep=comma]{data/sb-tlb.csv};

\node[] at (axis cs: 17, 3) {\scriptsize{TLB hit}};
\node[] at (axis cs: 3.5, 3) {\scriptsize{unmapped}};
\end{axis}
\end{tikzpicture}

%% file: img/spec-leak.tikz
\begin{tikzpicture}[yscale=.75]

\node[right] at (-0.5,-1.75) {\scriptsize 256 pages kernel memory (\texttt{kernel})};
\draw[step=.5] (0,-2.5) grid +(5.5,.5);
\node at (2.75,-2.25) {\small $\cdots$};

\node[right] at (0,-0.5) {\texttt{\textbf{if} (x < len(array))}};
\node[right] at (0,-1) {\qquad \texttt{y = kernel[array[x] * 4096]}};

\draw[thick] (5.25,-1) -| (5.5,-1.5) -- (4.75,-1.5) edge[->,>=stealth] node[right] {\small Store in TLB} (4.75,-2);

\draw[fill=yellow!30] (4.5,-2.5) rectangle (5,-2) node[midway] {\tiny \parbox{1cm}{\centering TLB\\Hit}};

\foreach \x in {1,...,8}
{
  \draw[densely dotted] (\x / 2 + .25, -3) -- (\x / 2 - 0.25, -3);
}

\foreach \x in {0,...,8,10}
{
    \draw[fill=blue] (\x / 2 + 0.25,-3) circle (.05);
}
\draw[fill=blue] (4.75,-3.25) circle (.05);
\draw[densely dotted] (4.75, -3.25) -- (5.25, -3);
\draw[densely dotted] (4.25, -3) -- (4.75, -3.25);
\node at (3,-3.3) {\small \AttackPrimitive};

\draw[densely dotted,thick] (-1,-2.75) -- (7,-2.75);
\node[right,black!70] at (-1,-2.5) {\small Kernel};
\node[right,black!70] at (-1,-3) {\small User};

\end{tikzpicture}

%% file: img/js_timings.tikz
\begin{tikzpicture}
\begin{axis}[
style={font=\footnotesize},
xlabel={Page},
ylabel={Increments},
width=0.95\hsize,
xmin=-0.2,
xmax=12.5,
scaled y ticks=false,
height=3cm
]
\draw[fill=black, opacity=0.2,draw=none] (axis cs: -0.5,-300) rectangle (axis cs: 9.5,6700);

\addplot+[densely dotted,mark=*,mark size=1.5pt] table[x=page,y=time,col sep=comma]{data/js_timings.csv}; 

\node[] at (axis cs: 3.5, 2500) {\scriptsize{unmapped}};
\end{axis}
\end{tikzpicture}

%% file: img/usbhid_access_trace.tikz
\begin{tikzpicture}
\begin{axis}[
style={font=\footnotesize},
xlabel={Sampling Period},
ylabel={Detection},
width=0.95\hsize,
xmin=0,
xmax=150,
ymin=0,
ymax=0.9,
scaled y ticks=false,
height=3cm,
legend pos=north west,
legend style={draw=none,fill=none}
]
\draw[fill=black, opacity=0.2,draw=none] (axis cs: 25,-2) rectangle (axis cs: 78,2);
\draw[fill=black, opacity=0.2,draw=none] (axis cs: 85,-2) rectangle (axis cs: 100,2);
\draw[fill=black, opacity=0.2,draw=none] (axis cs: 104,-2) rectangle (axis cs: 145,2);

\addplot+[only marks,mark=x,mark size=0.95pt] table[x=sample,y=hits,col sep=comma]{data/usbhid_access_trace.csv};
\addplot+[only marks,mark=diamond*,mark size=0.95pt] table[x=sample,y=ref,col sep=comma]{data/usbhid_access_trace.csv};

\addlegendentry{\scriptsize{target}}
\addlegendentry{\scriptsize{reference}}

\node[] at (axis cs: 51.5, 0.7) {\scriptsize{mouse movement}};
\node[] at (axis cs: 92.5, 0.7) {\scriptsize{keyboard}};
\node[] at (axis cs: 124, 0.7) {\scriptsize{mouse movement}};

\end{axis}
\end{tikzpicture}

%% file: img/bt_access_trace_hits.tikz
\begin{tikzpicture}
\begin{axis}[
style={font=\footnotesize},
xlabel={Sampling Period},
ylabel={hits [\%]},
width=0.95\hsize,
xmin=50,
xmax=250,
ymin=0,
ymax=1.4,
scaled y ticks=false,
height=3cm,
xtick={50,100,...,250},
xticklabels={0,50,...,200},
legend pos=south west,
legend style={draw=none,fill=none}
]
\draw[fill=black, opacity=0.2,draw=none] (axis cs: 92,-2) rectangle (axis cs: 139,2);
\draw[fill=black, opacity=0.2,draw=none] (axis cs: 149,-2) rectangle (axis cs: 154,2);
\draw[fill=black, opacity=0.2,draw=none] (axis cs: 166,-2) rectangle (axis cs: 194,2);
\draw[fill=black, opacity=0.2,draw=none] (axis cs: 207,-2) rectangle (axis cs: 233,2);

\addplot+[only marks,mark=x,mark size=0.95pt] table[x=sample,y=hits,col sep=comma]{data/bt_access_trace.csv};
\addplot+[only marks,mark=diamond*,mark size=0.95pt] table[x=sample,y=ref,col sep=comma]{data/bt_access_trace.csv};

\addlegendentry{\scriptsize{target}}
\addlegendentry{\scriptsize{reference}}

\node[] at (axis cs: 70, 1.2) {\scriptsize{quiet}};
\node[] at (axis cs: 115, 1.2) {\scriptsize{noise}};
\node[] at (axis cs: 153, 1.2) {\scriptsize{connect}};
\node[] at (axis cs: 180, 1.2) {\scriptsize{audio}};
\node[] at (axis cs: 220, 1.2) {\scriptsize{audio}};

\end{axis}
\end{tikzpicture}

%% file: img/bt_access_trace_detection.tikz
\begin{tikzpicture}
\begin{axis}[
style={font=\footnotesize},
xlabel={Sampling Period},
ylabel={Detection},
width=0.95\hsize,
xmin=50,
xmax=250,
ymin=0,
ymax=1.4,
scaled y ticks=false,
height=3cm,
xtick={50,100,...,250},
xticklabels={0,50,...,200}
]
\draw[fill=black, opacity=0.2,draw=none] (axis cs: 92,-2) rectangle (axis cs: 139,2);
\draw[fill=black, opacity=0.2,draw=none] (axis cs: 149,-2) rectangle (axis cs: 154,2);
\draw[fill=black, opacity=0.2,draw=none] (axis cs: 166,-2) rectangle (axis cs: 194,2);
\draw[fill=black, opacity=0.2,draw=none] (axis cs: 207,-2) rectangle (axis cs: 233,2);s

\addplot+[densely dotted,mark=x,mark size=1.0pt] table[x=sample,y=detected,col sep=comma]{data/bt_access_trace.csv};

\node[] at (axis cs: 70, 1.2) {\scriptsize{quiet}};
\node[] at (axis cs: 115, 1.2) {\scriptsize{noise}};
\node[] at (axis cs: 153, 1.2) {\scriptsize{connect}};
\node[] at (axis cs: 180, 1.2) {\scriptsize{audio}};
\node[] at (axis cs: 220, 1.2) {\scriptsize{audio}};

\end{axis}
\end{tikzpicture}